# Detection of acoustic phonons in carbon by Raman spectroscopy

Konstantin Iakoubovskii[1,*], Andrey Katrusha[2], Weihua Peng[2], Jianguo Peng[2]

[1] *Central European Institute of Technology, Brno University of Technology, Purkyňova 123, 61200 Brno, Czech Republic*
[2] *Jilin Diamond Technology Research & Development, Jianan Road 177, Luyuan District, Changchun, China*

**ABSTRACT**

*We detected acoustic phonons in graphite and diamond by Raman spectroscopy supported by density functional theory calculations. The activation of these normally-forbidden Raman modes was achieved via lattice amorphization in case of graphite and by boron doping in case of diamond. The doping-induced Raman signal in diamond was identified with substitutional boron of tetrahedral symmetry via its dependences on excitation wavelength and polarization. Comparison of the Raman spectra of amorphized graphite and heavily boron-doped diamond suggests the emergence of graphitic-like disorder in the diamond lattice. The reported approach is not limited to carbon and can be extended to a wide range of other materials.*

*Corresponding author: iakoubovskii@vut.cz

Acoustic phonons govern essential physical phenomena in solids, including thermal conductivity, elasticity, specific heat, propagation of sound and thermal expansion, yet their spectral characterization involves complex momentum-resolved techniques such as inelastic scattering of X-rays or neutrons. These techniques require relatively large and thick samples (ca. 10×10×1 mm) and hence are often inapplicable to novel and technologically important nanomaterials. Hence it would be highly desirable to probe acoustic phonons by optical methods like Raman scattering, which are widely accessible, easy to use, and can easily detect signals from nanosized volumes. However, such measurements are hindered by momentum and symmetry related selection rules for optical transitions that strictly apply to most crystals – specifically, to centrosymmetric crystals. This is because acoustic phonons correspond to uniform translation of the whole lattice. Hence, they do not change polarizability and induce no Raman signals in first order.

This problem had been addressed by relaxing the lattice-induced selection rules back in 1975 by Alben *et al.*,[1] who characterized acoustic modes in Si and Ge crystals by studying Raman scattering from their amorphous counterparts. More recently, acoustic modes have been detected in carbon nanostructures via double-resonance Raman scattering;[2,3] however, this technique is more indirect and requires detailed analysis of weak overtone peaks.

The amorphization method of Alben *et al.* yielded interesting new data, but their analysis was hindered by the need to use a series of crystalline and amorphous samples prepared by vastly different techniques. Later it was extended from Si and Ge to some other materials such as GaAs[4] and ZnO[5], but not to carbon. The main difficulty with carbon is due to its rich variety of phases that are hard to control in a disordered state. Particularly difficult to avoid are graphitic forms, which are more stable than diamond under ambient or near-ambient conditions. As a result, numerous attempts to amorphize diamond or prepare 100% $sp^3$-hybridized carbon always resulted in prominent $sp^2$-carbon-related Raman signals.[6]

In this work we report detection and characterization of acoustic phonons by Raman scattering in two major forms of carbon, diamond and graphite. For graphite, we refined the approach of Alben *et al.*[1] and recorded Raman signals from the same sample before and after its amorphization. For diamond, we activated a Raman signal from acoustic phonons via boron doping, which broke the local lattice symmetry. We identified that signal with substitutional boron acceptor via its dependences on polarization and excitation wavelength.

Boron-doped diamond crystals were synthesized by the standard high-pressure high-temperature (HPHT) technique using a cubic press, a Co-Fe-C growth system, and a proprietary nitrogen getter. We used a pressure of 6.0–6.5 GPa and a temperature of 1600–1650 ºC. Light doping was achieved by introducing 0.5–1 wt% of amorphous boron to the growth mixture, which resulted in the lattice boron concentration of ca. 0.01 at%, as evaluated via Fourier-transform infrared absorption (Vertex70v, Bruker). The as-grown samples were thoroughly cleaned with aqua regia, hydrochloric acid, acetone and water.

Heavily-doped diamond (1–2 at%) was purchased from Bruker. It was grown by chemical vapor deposition (CVD) as ca. 100 nm thick films on silicon. Commercial pyrolytic graphite was amorphized by irradiation with 30 keV $Xe^+$ ions inside an Amber X2 scanning electron microscope – focused ion beam system (SEM-FIB, Tescan) at a dose of $10^{18}$ ions/cm$^2$.

All measurements were performed at room temperature. Nearly all Raman spectra were recorded in backscattering geometry with WITec alpha 300R confocal microscopy setup equipped with 633 nm (HeNe), 355 and 532 nm (Nd:YAG) lasers. To extend Raman measurements into the infrared range, we employed a Bruker RFS 100 Fourier-transform spectrometer equipped with a 1064 nm Nd:YAG laser.

Morphology and chemical composition of the samples was characterized with the Amber X2 system, which besides SEM-FIB was equipped with an energy-dispersive X-ray (EDS) and time-of-flight secondary ion mass (ToF-SIMS) spectrometers. Chemical composition was also measured by X-ray photoelectron spectroscopy (XPS, Kratos Axis Supra). XPS confirmed that the $sp^2/sp^3$ ratio in graphite is close to 100%, both before and after $Xe^+$ irradiation.

Density functional theory (DFT) calculations of phonon dispersions and phonon density of states (PDOS) were performed on ideal diamond and graphite lattice structures using the Quantum Espresso 7.5 package. We applied a norm-conserving pseudopotential generated within the Perdew–Burke–Ernzerhof (PBE) formulation of the generalized gradient approximation (GGA) exchange–correlation functional, an 8×8×8 supercell, a wavefunction cut-off of 100 Ryd, and a 20×20×20 mesh in k-space for diamond. For graphite, we adopted a 6×6×3 supercell and 24×24×12 k-mesh, and its amorphization was considered by introducing Gaussian broadening into calculations. This approach is overly simplistic, yet it is cost-effective and adequate for this study that does not aim at detailed modeling of amorphization effects.

Fig. 1a shows characteristic Raman spectra of lightly (ca. 0.01 at.%, HPHT) and heavily boron-doped diamond (ca. 1 at.%, CVD). The spectrum is dominated by the 1332 $cm^{-1}$ signal that corresponds to a lattice optical phonon of tetrahedral symmetry. Boron doping induces three peaks at ca. 590, 900 and 1050 $cm^{-1}$, which usually appear together.[7] However, our measurements at different sample depths (Fig. 1a) reveal that they have dissimilar spatial distributions (and polarization dependences) and hence belong to different defects. Furthermore, the 900 and 1050 $cm^{-1}$ features disappear at high doping level (see top curve in Fig. 1a), and hence we focus on the 590 $cm^{-1}$ peak.

Fig. 1b compares the variation of the 590 $cm^{-1}$ peak intensity with the excitation wavelength $\lambda$ (blue squares) and the optical absorption spectrum (black line), which is associated with the boron acceptor in diamond.[8] A close resemblance is observed, suggesting that light absorption at the boron acceptor is responsible for the 590 $cm^{-1}$ peak. Note that the observed strong increase in Raman intensity I with $\lambda$ is unusual, and a $I \sim \lambda^{-4}$ behavior is much more common.

Polarization dependences of Fig. 1c,d reveal that the 590 $cm^{-1}$ peak preserves the tetrahedral symmetry of the diamond lattice,[7] supporting the above assignment. Please note that the polarization dependences of Fig. 1c,d were measured by rotating the analyzer rather than diamond sample. This greatly reduces experimental errors that inevitably arise when rotating a sample under a confocal microscope, while the analysis of results is similar and straightforward.[9]

Another important note is that the experimentally supported assignment of the 590 $cm^{-1}$ peak to a tetrahedral defect and substitutional boron in this article is not trivial, as this peak was widely speculated to originate from boron-boron pairs instead.[10] The tetrahedral symmetry of the phonon responsible for 590 $cm^{-1}$ peak strongly suggests that this phonon originates from a carbon lattice vibration. Indeed, the position of this peak differs in boron-doped $^{12}C$ and $^{13}C$ diamonds, while no shift is observed when $^{10}B$ is replaced with $^{11}B$ isotope.[11] To further elaborate its nature, we compared Raman spectra with DFT calculations.

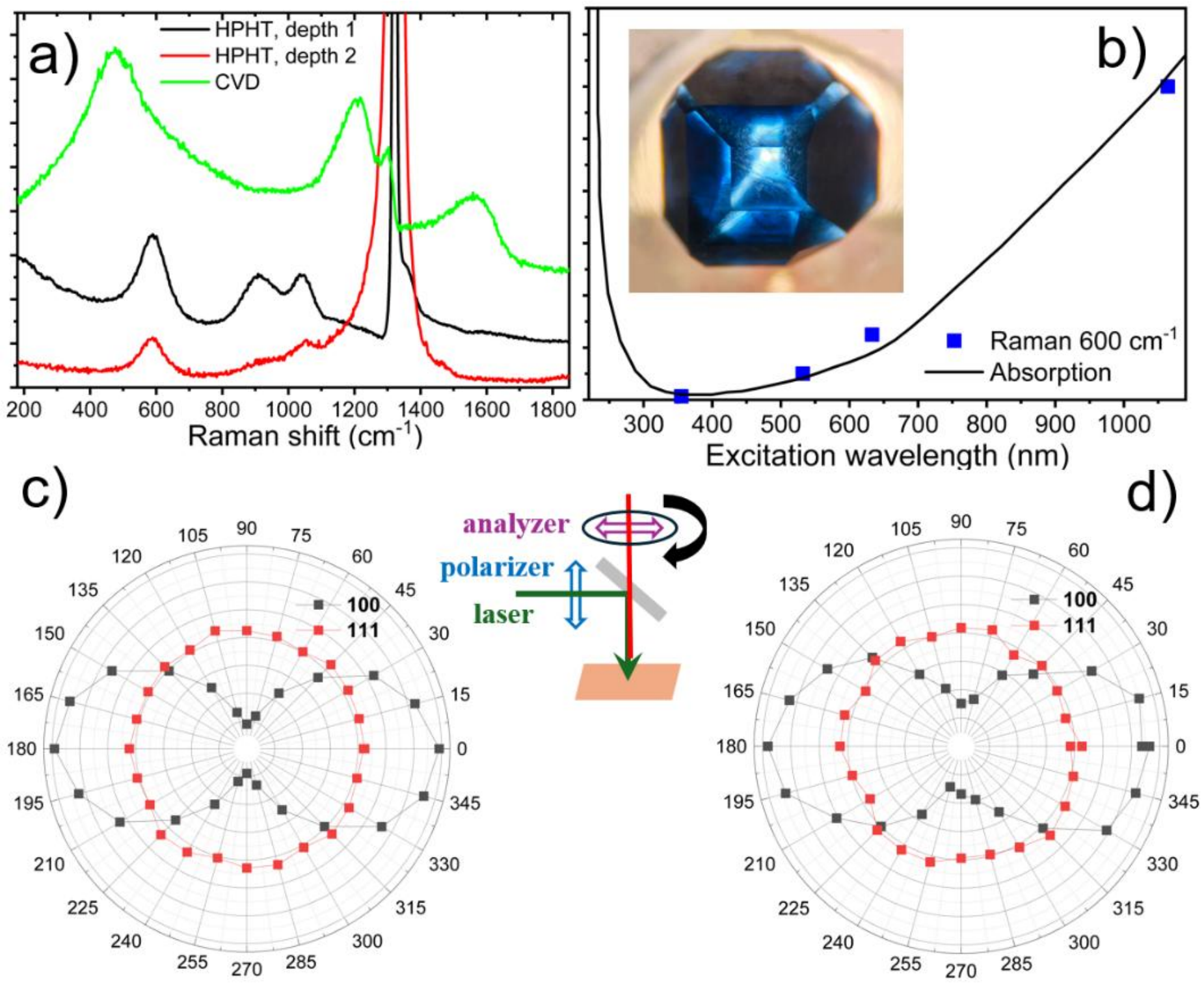

*Fig. 1. a) Raman spectra measured at two different depths of a lightly boron doped HPHT diamond (ca. 0.01 at%, see inset in panel b) under 532 nm excitation. The top curve shows a spectrum from heavily doped (ca. 1 at%) CVD diamond film. b) Variation of the 590 $cm^{-1}$ Raman peak intensity with the excitation wavelength (solid blue squares) compared with the optical absorption spectrum. c,d) Polarization dependences of the 1332 $cm^{-1}$ (c) and 590 $cm^{-1}$ (d) peak intensity vs. the angle between polarizer and analyzer for light incident on the (100) and (111) diamond planes. The inset outlines polarization geometry.*

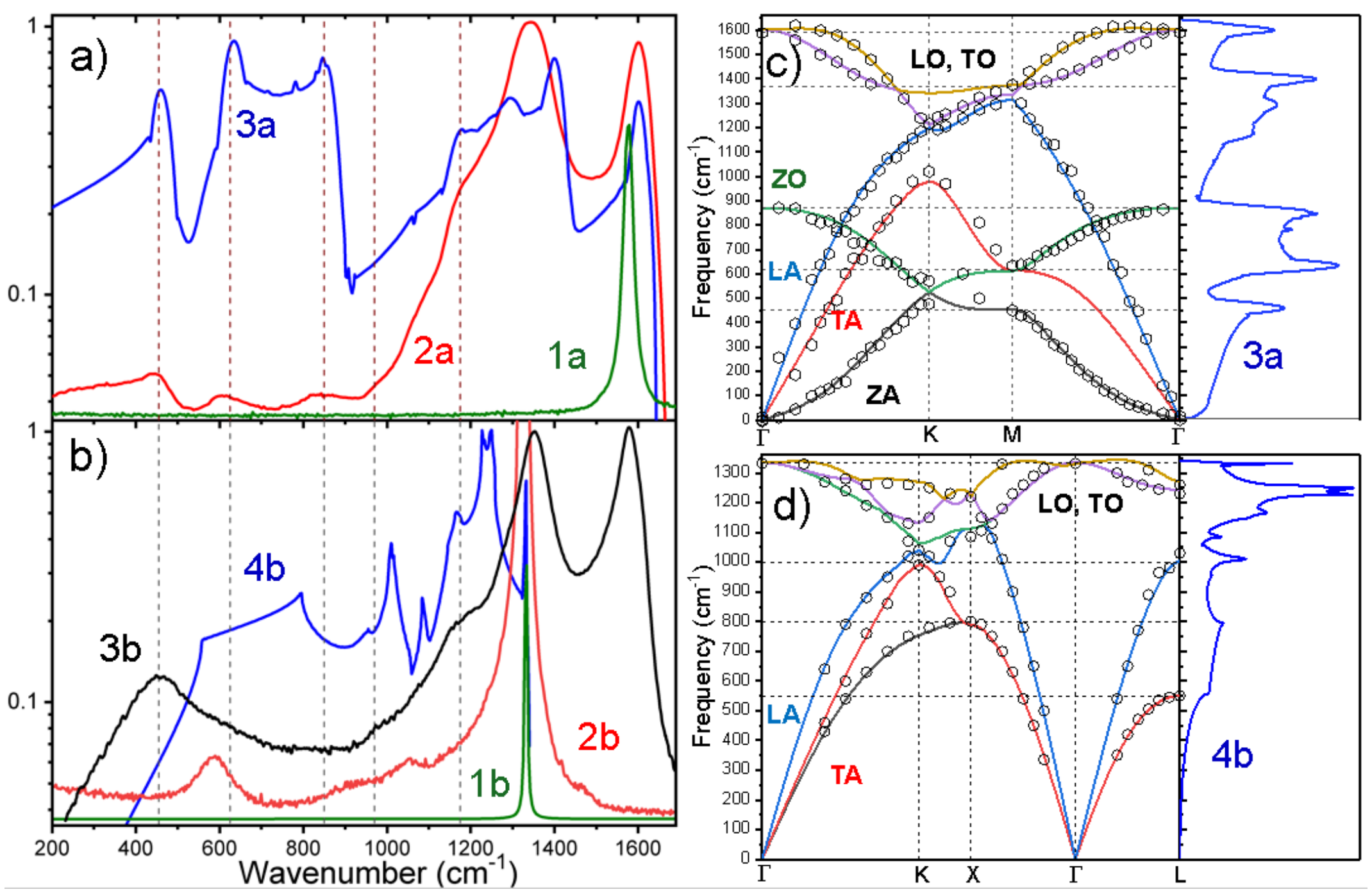

*Fig. 2. Experimental Raman spectra (a,b, 532 nm excitation) and calculated phonon dispersion branches (c,d) for partly amorphized graphite (a,c) and crystalline diamond (b,d). Note the semilogarithmic scale in (a,b). The phonon dispersions were calculated for perfect crystals. Open symbols in (c,d) present experimental data compiled from the literature.[12] Curves: 1a and 1b – Raman spectra of undoped diamond and graphite, respectively; 2a– Raman spectrum of amorphized graphite; 3a and 4b – PDOS of amorphized graphite and crystalline diamond, respectively, the amorphization was modeled by Gaussian broadening of the crystalline PDOS; 2b and 3b– Raman spectra of lightly (ca. 0.01 at.%) and heavily boron-doped diamond (ca. 2 at%), respectively. Phonon labels: LO/LA – longitudinal optical/acoustic, TO/TA – transverse optical/acoustic, ZO/ZA – out-of-plane (z-axis) optical/acoustic.*

Fig. 2a-b presents a summary of Raman results plotted in a semilogarithmic scale for adequate presentation of weak signals. As expected for perfect crystals, curves 1a and 1b, which represent graphite and diamond, respectively, show a single peak that corresponds to transverse optical (TO) and longitudinal optical (LO) phonons co-existing at the Γ point in the phonon dispersion diagrams (Fig. 2c-d).

Amorphization of graphite (curve 2a) results in the broadening and blue shift of the initially present LO/TO peak at ca. 1580 cm$^{-1}$, which is traditionally called G peak in the literature, and the appearance of a prominent defect-related D band at ca. 1350 cm$^{-1}$. Furthermore, weaker peaks can be noticed at ca. 460, 620 and 840 cm$^{-1}$.

The 460, 620 and 840 cm$^{-1}$ peaks in curve 2a match reasonably well the kinks in the calculated PDOS shown by curve 3a. Using the associated phonon dispersion curves (Fig. 2c), they can be tentatively assigned to ZA, TA and ZO phonons, respectively on the basis of their correspondence with features in the calculated PDOS.

When comparing the Raman and PDOS spectra, we would note in passing that some PDOS modes can be active only in infrared, but not in Raman spectra.

Light boron doping of diamond (ca. 0.01 at.%, see curve 2b in Fig. 2b) induces a low-frequency Raman peak. By comparing its position at ca. 590 cm$^{-1}$ to the calculated PDOS (curve 4b), this peak can be assigned to the TA mode at L-point.

Please note that this assignment apparently contradicts to the strong polarization dependences in the (100) plane presented in Fig. 1c,d. Indeed, averaging over four equivalent L-valleys, which individually have a trigonal symmetry, should flatten polarization dependences both for the (100) and (111) plane. This apparent contradiction can be rationalized as follows: substitutional boron merely relaxes the translational selection rule, enabling first-order scattering by finite-wavevector phonons, while preserving the local tetrahedral symmetry of the diamond lattice. Consequently, the effective Raman scattering operator is the same for the 590 and 1332 cm$^{-1}$ peaks, even though they originate at different symmetry points of the Brillouin zone (L and Γ).

The 590 cm$^{-1}$ peak shifts with boron doping down to 450 cm$^{-1}$ at 2 at%, as shown by curve 3b in Fig. 2b. This shift is accompanied by progressive disordering of the diamond lattice, as suggested by the transformation of the 1332 cm$^{-1}$ diamond Raman peak into the D and G bands characteristic of disordered carbon (compare curves 2a and 3b in Fig. 2a,b and note that this transformation is well documented in the literature[13]). Actually, the final position of the boron-induced peak in *diamond* matches that of ZA phonon in amorphous *graphite*.

Note that boron doping at 2 at% increases the diamond lattice constant $a$ by ca. 0.33%,[14] and the shift $\Delta\nu$ in the Raman frequency $\nu$ upon this lattice expansion can be evaluated as follows:

$$\frac{\Delta\nu}{\nu} = -\gamma\frac{\Delta V}{V} \approx -3\gamma\frac{\Delta a}{a} \quad (1)$$

Here V is volume and $\gamma = 0.96$ is the Grüneisen constant.[15] This estimate yields $\Delta\nu/\nu \sim 0.01$, which is much smaller than the $\Delta\nu/\nu \sim 0.23$ change that we observe for the 590 $cm^{-1}$ peak upon 2 at% boron doping. Hence, while lattice expansion upon boron doping must downshift the boron-induced Raman peak, its contribution should be minor.

Our conclusion that heavy boron doping may introduce substantial local disorder brings ambiguity to the common practice where the boron concentration in diamond is determined via the position of the boron-induced Raman peak.[15] Another implication of the doping-induced amorphization is potential inactivation of the boron acceptors, which may explain the saturation of electrical conductivity in boron-doped diamond at concentrations above 1 at.%.[16]

In summary, we demonstrated two methods of detecting normally-forbidden Raman phonon modes by breaking the inversion symmetry of the host crystal: introducing disorder via ion irradiation and doping the crystal. As a result, we assigned low-frequency Raman peaks in disordered graphite, which were observed before but left undiscussed,[17] to the ZA and TA modes at ca. 460 and 620 $cm^{-1}$, respectively. We have also revealed that the 590 $cm^{-1}$ mode, which is commonly observed in low-boron diamond,[7] is consistent with scattering involving the TA phonon near the L-point in the phonon dispersion. This mode red-shifts upon boron doping down to frequently quoted number of 500 $cm^{-1}$ or even below. Yet, this downshifted peak may involve both the TA mode and disorder, especially at heavy doping (>1 at.%). Disorder may result in electrical deactivation of boron acceptors and should be considered when evaluating the boron concentration in diamond by Raman spectroscopy.

## ACKNOWLEDGMENTS

CzechNanoLab project LM2023051 funded by MEYS CR is gratefully acknowledged for the financial support of the measurements at CEITEC Nano Research Infrastructure.

## CONFLICT OF INTEREST

The authors declare no conflicts of interest.